\documentclass[12pt,preprint]{aastex}
\usepackage{natbib}
\usepackage{amssymb,amsmath}

\shorttitle{ GRB 110530A }
\shortauthors{Zhong et al. }

\begin{document}

\title{GRB 110530A: Peculiar Broad Bump and Delayed Plateau in Early Optical Afterglows}

\author{Shu-Qing Zhong \altaffilmark{1, 2}, Li-Ping Xin \altaffilmark{3}, En-Wei Liang \altaffilmark{1, 2, 3, 4}, Jian-Yan Wei \altaffilmark{3}, Yuji Urata \altaffilmark{5,6}, Kui-Yun Huang \altaffilmark{7}, Yu-Lei Qiu \altaffilmark{3}, Can-Min Deng \altaffilmark{1, 2}, Yuan-Zhu Wang \altaffilmark{4}, Jin-Song Deng \altaffilmark{3}}

\altaffiltext{1}{GXU-NAOC Center for Astrophysics and Space Sciences, Department of Physics, Guangxi University, Nanning 530004, China; lew@gxu.edu.cn}

\altaffiltext{2}{Guangxi Key Laboratory for the Relativistic Astrophysics, Nanning 530004, China}

\altaffiltext{3}{Key Laboratory of Space Astronomy and Technology, National Astronomical Observatories, Chinese Academy of Sciences, Beijing 100012, China; xlp@bao.ac.cn}

\altaffiltext{4}{Purple Mountain Observatory, Chinese Academy of Sciences, Nanjing 210008, China}

\altaffiltext{5}{Institute of Astronomy, National Central University, Chung-Li 32054, Taiwan}

\altaffiltext{6}{Academia Sinica Institute of Astronomy and Astrophysics, Taipei 106, Taiwan}

\altaffiltext{7}{Department of Mathematics and Science, National Taiwan Normal University, Lin-kou District, New Taipei City 24449, Taiwan}

\begin{abstract}
We report our very early optical observations of GRB 110530A and investigate its jet properties together with its X-ray afterglow data. A peculiar broad onset bump followed by a plateau is observed in its early R band afterglow lightcurve. The optical data in the other bands and the X-ray data are well consistent with the temporal feature of the R band lightcurve. Our joint spectral fits of the optical and X-ray data show that they are in the same regime, with a photon index of $\sim 1.70$. The optical and X-ray afterglow lightcurves are well fitted with the standard external shock model by considering a delayed energy injection component. Based on our modeling results, we find that the radiative efficiency of the GRB jet is $\sim 1\%$ and the magnetization parameter of the afterglow jet is $<0.04$ with the derived extremely low $\epsilon_B$ (the fraction of shock energy to magnetic field) of $(1.64\pm 0.25)\times 10^{-6}$. These results indicate that the jet may be matter dominated. Discussion on delayed energy injection from accretion of late fall-back material of its pre-supernova star is also presented.
\end{abstract}

\section{Introduction}
It is generally believed that cosmic gamma-ray bursts (GRBs) are from ultra relativistic jets powered by newly-born black holes or pulsars during collapses of massive stars or mergers of compact stars (e.g., Colgate 1974, Paczynski 1986; Eichler et al. 1989; Narayan et al. 1992; Woosley 1993; MacFadyen \& Woosley 1999; Zhang et al. 2003; see reviews by M\'{e}sz\'{a}ros 2002, 2006; Zhang \& M\'{e}sz\'{a}ros 2004; Piran 2004; Woosley \& Bloom 2006; Kumar \& Zhang 2015). Their prompt gamma-ray emission may be from internal shocks in an erratic, unsteady, relativistic fireball (e.g., Rees \& M\'{e}sz\'{a}ros 1992; M\'esz\'aros \& Rees 1993; Rees \& M\'{e}sz\'{a}ros 1994), a dissipative
photosphere ({e.g., Beloborodov 2010; Vurm et al. 2011; Giannios 2008; Ioka 2010}), or a Poynting-flux-dominated outflow ({ Zhang \& Yan 2011 and reference therein}). The broad band observations with the Fermi mission sharpen debating on the radiation mechanisms and the composition of the GRB jets (e.g., Abdo et al. 2009; Zhang et al. 2009, 2011; Zhang et al. 2013; Lyu et al. 2014).

Long-lived afterglows in the X-ray, optical and radio bands following the prompt gamma-rays were discovered in the BeppoSAX mission era (van Paradijs et al. 2000 and references therein). They are well explained with the synchrotron emission from external shocks when GRB fireballs propagate into the circumburst medium (e.g., M\'esz\'aros \& Rees 1997; Sari et al. 1998). Afterglow observations was revolutionized by the {\em Swift} mission thanks to the promptly slewing and precisely localizing capacities of its X-ray telescope (XRT) (Gehrels et al. 2004; Burrows et al. 2005b). The number of GRBs that have optical and X-ray afterglow detections rapidly increases and the sample of well-sampled lightcurves are also growing quickly (Gehrels et al. 2009; Kann et al. 2010). Excluding the tail emission of the prompt gamma-rays and erratic flares from the canonical XRT lightcurves (Nousek et al. 2006; Zhang et al. 2006), the X-ray afterglow lightcurves are generally consistent with the predictions of the external shock model by adding an extra energy injection (Zhang et al. 2006; Liang et al. 2007). Statistical analysis of the optical afterglow lightcurves observed from Feb, 1997 to Nov., 2011 shows that about 1/3 of the optical afterglow lightcurves well agree with the prediction of the external shock model in the thin shell case, and another 1/3 may require an extra energy injection to the external shocked medium (Li et al. 2012; Liang et al. 2013). An extensive analysis of the X-ray and optical afterglow data by Wang et al. (2015) shows that the standard external shock models are good for explaining the data by elaborately considering various effects, such as long-lasting reverse shock, structured jets, circumburst medium density profile.

Well-sampled multi-wavelength lightcurves in broad temporal coverage from very early to late epochs are valuable for modeling the lightcurves and revealing the properties of the GRB jets and even the GRB central engines as well as the progenitors (e.g., Xin et al. 2016). This paper reports our very early optical observations for GRB 110530A and detailed modeling for the optical and X-ray afterglow lightcurves. Observations and data reductions are reported in \S 2. We present joint temporal and spectral analysis for the optical and X-ray afterglow data in \S 3, and present our modeling results in \S 4. Discussion on the possible implications for its jet composition and progenitor star are available in \S 5. Conclusions are presented in \S 6. Throughout, the notation $Q_n=Q/10^{n}$ in cgs units are adopted.

%
%
%
%
%
%
%
%
%

\section{Observations and Data Reduction}
XRT and UV-optical Telescope (UVOT) on board {\em Swift} began observed the X-ray and optical afterglows of GRB 110530A at 446 seconds and 438 seconds after the {\em Swift} Burst Alert Telescope (BAT) trigger, respectively (D'Avanzo et al. 2011a, b). Our optical follow-up observations began much earlier than the first detections of XRT and UVOT (Marshall et al. 2011). The TNT (0.8-m Tsinghua University - National Astronomical Observatory of China
Telescope) at Xinglong Observatory\footnote{TNT is a 0.8-m telescope and runs by a custom-designed automation system for GRB follow-up observations at Xinglong Observatory. A PI $1300\times1340$ CCD and filters in the standard Johnson Bessel system are equipped for TNT (Zheng et al. 2008).} promptly slewed to the burst position 133 seconds after the {\em Swift}/BAT trigger, and the optical counterpart was clearly detected in all images in the $white (W)$ and $R$ bands. The early optical afterglows of GRB 110530A was also observed with AZT-33IK telescope of Sayan observatory (Mondy) and well-sampled lightcurve was obtained (Volnova et al. 2011). Our observations with Lulin One-meter Telescope ( LOT ) at Taiwan started at about 30 min after the burst, and the optical counterpart was also clearly detected in the $g$, $r$, and $i$ bands. The optical counterpart was also detected with the 2.5m NOT telescope at Roque de los Muchachos Observatory (La
Palma, Spain) at 6.8 hours after the burst. It faded down to $R\sim 21.3$ mag (De Cia et al. 2011). Spectroscopic
observations with NOT does not show any evident absorption lines, and a limit of the redshift $z<2.7$ is placed by the non-detection of Lyman alpha absorption in the spectra (De Cia et al. 2011). We assume that $z=1$ for our analysis.

\begin{figure}[htbp]
 \centering
\includegraphics[angle=0,width=0.45\textwidth]{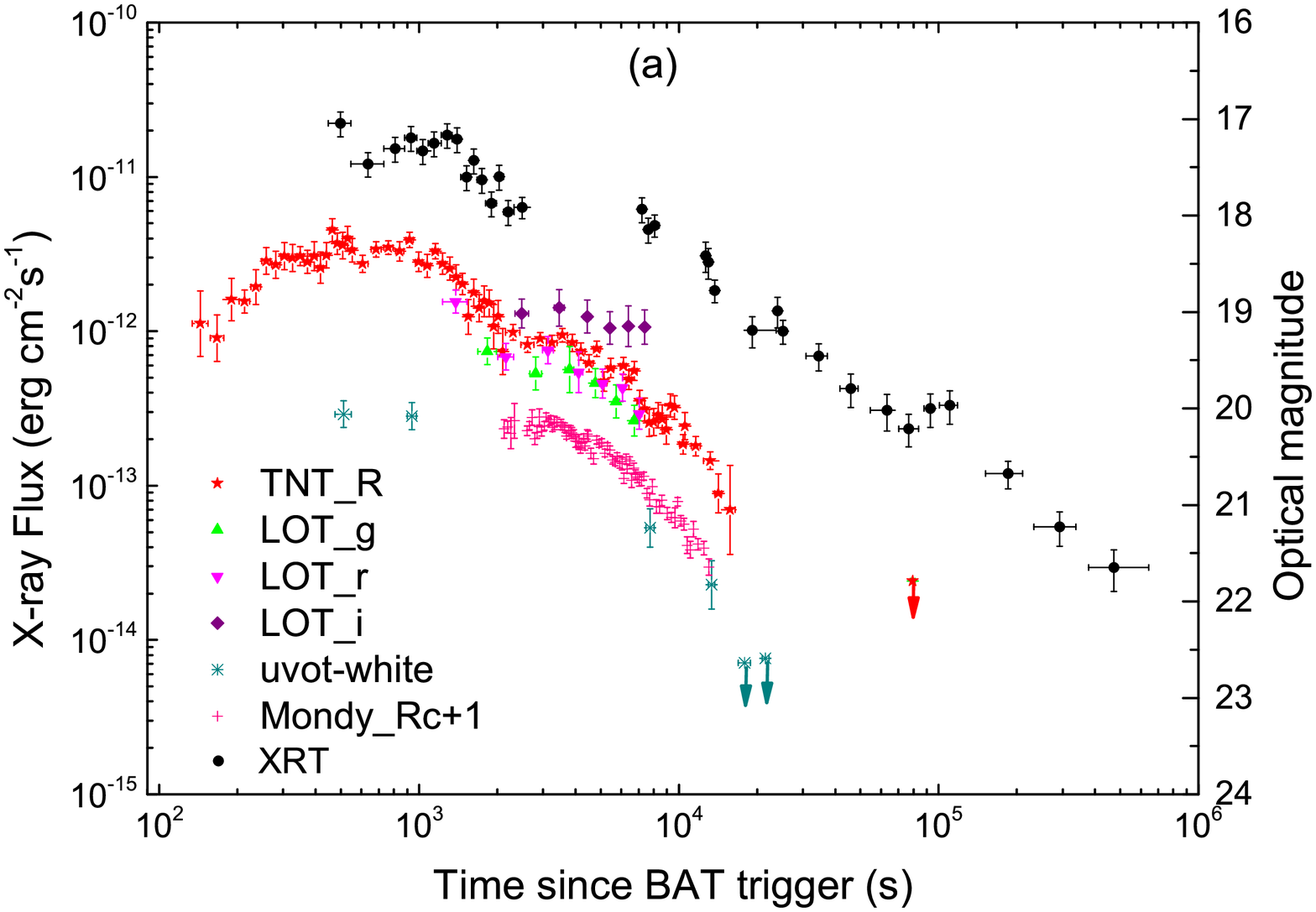}
\includegraphics[angle=0,width=0.45\textwidth]{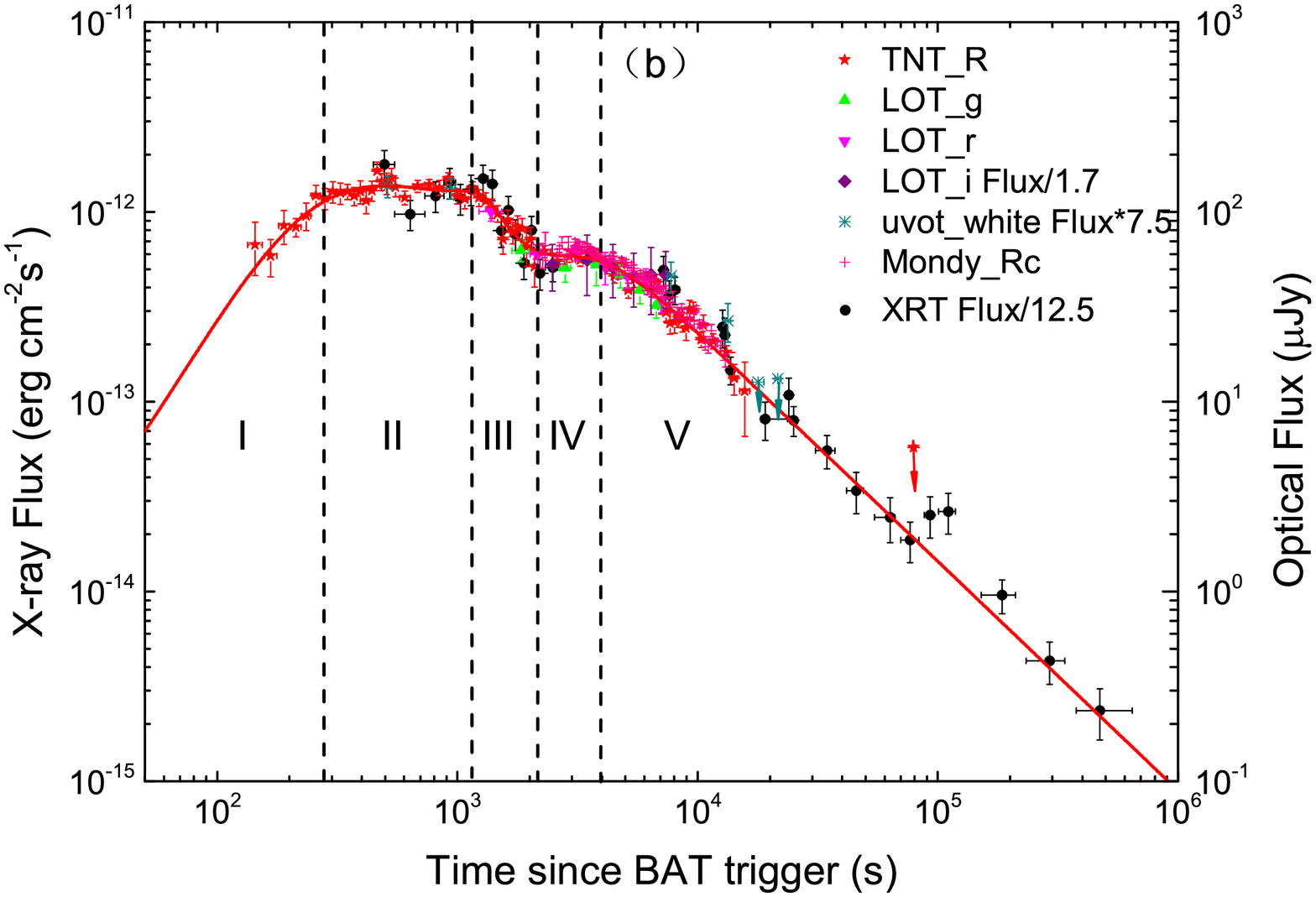}
\caption{Observed optical and X-ray afterglow lightcurves of GRB 110530A ({\em left panel}) and our empirical fit with multiple smooth broken power-laws for the R band lightcurves ({\em Right panel}). The optical data in the bands and XRT data in the Right panel are re-scaled in order to show their consistency of the temporal feature with the R band lightcurve. Phases identified from our empirical fit are also marked. The early optical afterglow data observed with AZT-33IK telescope of Sayan observatory (Mondy) read from Volnova et al. (2011) was also illustrated for comparison.}
\label{obs_lc}
\end{figure}

We process our optical data by following the standard routine in the IRAF  package\footnote{IRAF is distributed by NOAO, which is operated by AURA, Inc. under cooperative agreement with NSF.}. Point spread function (PSF) photometry was applied with the DAOPHOT tool in the IRAF package to obtain the instrumental magnitudes. For the $white$ band data, we simply take them as $R$ band data (Xin et al. 2010).  All TNT optical data were calibrated by USNO B1.0 R2 mag with 11 nearby reference stars. The  data observed with the LOT telescope was calibrated with the transformation of Jordi et al. (2006)\footnote{http://classic.sdss.org/dr6/algorithms/sdssUBVRITransform.html\#Jordi2006} with USNO B1.0 mag.
Our optical observations are reported in Table 1 and the optical afterglow lightcurves are shown in Figure \ref{obs_lc}.
The reference stars for calibration is presented in Table 2.

The {\em Swift}/XRT lightcurve and spectrum are extracted from the UK Swift Science Data Centre at the University of Leicester (Evans et al. 2009)\footnote{http://www.swift.ac.uk/results.shtml}. The XRT lightcurve with 30 counts per bin is also shown in Figure \ref{obs_lc}.

The duration of prompt emission in the BAT band is $T_{\rm 90}= 19.6$ s. we extract the prompt gamma-rays spectrum following the standard BAT data processing routine. It is well known the GRB spectrum in the keV-MeV band is empirically fit by the Band function with typical photon indices $\Gamma_1=-1$ and $\Gamma_2=-2.3$ breaking at $E_{\rm b}$ (Band et al. 1993; Preece et al. 2000). The peak energy of the $\nu f_\nu$ spectrum is given by $E_{\rm p}=(1+\Gamma_1)$ if $\Gamma_2<-2$. $E_p$ value may vary from tens to thousands keVs among GRBs. Since BAT energy band is only 15-150 keV, the GRB spectrum observed with BAT is usually adequately fitted with a single power-law, and an empirical relation between $E_{\rm p}$ and $\Gamma_\gamma$ is proposed, i.e., $\log E_{\rm p}=(2.76\pm 0.07)-(3.61\pm 0.26)\log \Gamma$ (Zhang et al. 2007b). Fitting the BAT spectrum of GRB 110530A with a single power-law, we get $\Gamma_\gamma=2.04\pm 0.21$, and its fluence in BAT energy band is $3.3\times10^{-7}$erg cm$^{-2}$ in this spectral model. With the empirical relation between $E_{\rm p}$ and $\Gamma_\gamma$ , we have $E_{\rm p}\sim 45$ keV. Correcting the $E_{\rm \gamma, iso}$ in the BAT band to $1-10^4$ keV band with the spectral parameter $\Gamma_1=-1$, $\Gamma_2=-2.3$, and $E_{\rm p}=45$ keV, we obtain $E^{\rm c}_{\rm \gamma, iso}=1.92\times 10^{51}$ erg assuming $z=1$. With the spectral parameters, we also obtain the peak luminosity in the $1-10^4$ keV band as $L^{\rm c}_{\rm \gamma, iso}=(2.81\pm 0.71)\times 10^{50}$ erg s$^{-1}$.

\section{Data Analysis}
As shown in Figure \ref{obs_lc}, well-sampled lightcurve in the R band is observed with the TNT. We empirically fit the lightcurve with a multiple broken power-law model. Each broken power-law function is read as,
\begin{equation} F=F_0\left [
\left (   \frac{t}{t_b}\right)^{\omega\alpha_1}+\left (
\frac{t}{t_b}\right)^{\omega\alpha_2}\right]^{1/\omega},
\end{equation}
where $t_b$ is the break time,$\alpha_1$ and $\alpha_2$ are decay indices before and after the break, respectively, and $\omega$ describes the sharpness of the break. Our fit yields five phases, as shown in the right panel of Figure \ref{obs_lc}.
The R band lightcurve smoothly onsets with a slope of $2.6\pm 0.4$ (Phase I) and peaks at $275\pm 22$ seconds. The flux keeps almost constant from 275 seconds to 1300 seconds (the first plateau, Phase II), then decays with a power-law index of -1.2 (Phase III). Subsequently, the flux keeps almost constant (the second plateau, Phase IV) and transits to a normal decay with a power-law of -1.2 again (Phases V). Flickering is shown up at around $T_0+460$ and $T_0+1200$ seconds during the first optical plateau. By re-scaled the multi-band optical data and XRT data, it is clear shown that both other-wavelength optical data and X-ray data are well consistent with the temporal feature of the R band lightcurve, even the optical flickering feature is also clearly shown up in the X-ray band. These results confidently indicate that the optical and X-ray afterglows are from the same emission component. {Such a light curve shape has been seen before in other GRB afterglows, though with a less pronounced early plateau, such as GRB 071025 (Perley et al. 2010), GRB 091024 (Virgili et al. 2013), and GRB 110213A (Cucchiara et al. 2011). While it is lacking a second hump, an early rise-plateau-decay was also recently reported for GRB 141221A (Bardho et al. 2016).}

\begin{figure}[htbp]
 \centering
\includegraphics[angle=0,width=0.5\textwidth]{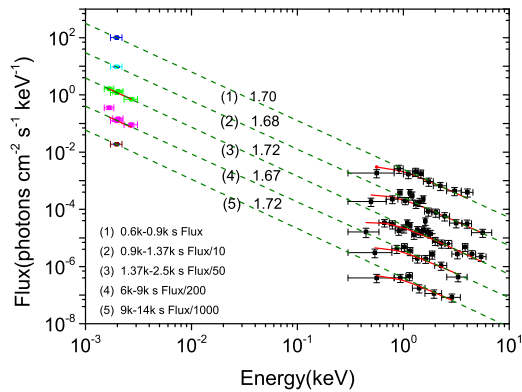}
\caption{Joint spectral fits for the optical and X-ray afterglows with a single power-law function in selected five time intervals. The {\em Olive dashed lines} shows that the intrinsic power-law spectrum derived from the joint fits. The photon indices are also marked.}
\label{obs_spec}
\end{figure}

To investigate the spectral properties of the afterglow data, we extract the joint optical and X-ray spectra of the afterglows in five time intervals, i.e., 0.6-0.9 ks, 0.9-1.37 ks, 1.37-2.5 ks, 6-9 ks and 9-14 ks. The X-ray data in each time intervals are grouped with a criterion of 10 counts per bin. The selected time intervals are for the Phase II-V and late epoch of Phase V. Spectral analysis for Phase I could no be made since no X-ray data is available. The optical data is corrected by the extinction of our Galaxy, which are $A_{g}=0.182$,$A_{r}=0.126$,$A_{R}=0.119$ and $A_{i}=0.093$ at the burst direction (Schlegel et al. 1998). The equivalent hydrogen column density of our Galaxy is $N_{\rm H}=6.78\times 10^{20}$ cm$^{-2}$. We use the Xspec package to analyze the spectral data. The extinction laws of host galaxy is taken as that of Large Magellanic Cloud (LMC; $R_{\rm V}=3.16$) and Small Magellanic Cloud (SMC; $R_{\rm V}=2.93$). The $N_{\rm H}$ of the host galaxy is derived from the time integrated X-ray afterglow spectrum. It is $N_{\rm H}^{\rm host}\sim 1.0\times 10^{21}$cm$^{-2}$, which is fixed at this value in our time-resolved spectral fits. Considering hydrogen absorptions and extinctions of both our Galaxy and host galaxy, we fit the spectra with a single power-law function. Our results are reported in Table 3 and shown in Figure \ref{obs_spec}. The derived photon indices range from 1.67 to 1.72. The extinction by the host galaxy is negligible for both the LMC and SMC extinction laws\footnote{ Note that the redshift of GRB 110530A is unknown and we have only an upper limit of $z<2.7$ (De Cia et al. 2011). Our dust modelings may be insecure since the LMC and SMC extinction curves, especially the LMC dust curve, have features which become relevant in this redshift range.}

\section{Modeling the optical and X-ray afterglow lightcurves}
Our temporal and spectral analysis shows that the optical and X-ray afterglows are from the same emission component. The clear detection of the smoothly onset feature in the early optical data is well consistent with the expectation of the standard external shock model in the thin shell case (Sari \& Piran. 1999; Liang et al. 2010, 2013). The observed first plateau seems to be shaped by the broadening of the onset bump with the superimposed flares (or flickering), which may be due to fluctuations of the external shock region or due to flares from late internal shocks (e.g., Burrows et al. 2005a; Fan et al. 2005; Zhang et al. 2006; Dai et al. 2006; Liang et al. 2006). We do not consider these erratic flares in our modeling. The second plateau from 2400 s to 3000 s could be attributed to delayed  energy injection to the afterglow jet. Therefore, we model the lightcurves with the standard afterglow models by considering the late energy injection effect. We adopt the standard afterglow model by Sari et al. (1998) and Huang et al. (1999). We describe our model fitting strategy as following.

\begin{figure}
 \centering
\includegraphics[angle=0,width=0.5\textwidth]{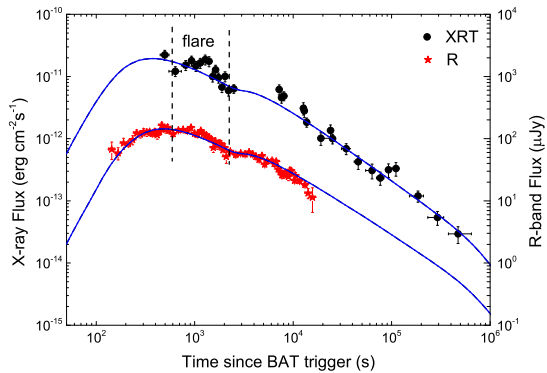}
\caption{Fits to the optical and X-ray afterglow lightcurves using the standard external shock model by considering a delayed energy injection behaving as $L_{\rm in}=L_0 t^{q}$. The model parameter derived from the MCMC technique are $\Gamma_0=91\pm8$, $\epsilon_e=0.086\pm0.008$, $\epsilon_B=(1.64\pm0.25)\times10^{-6}$, $n=13.3\pm2.6$ cm$^{-3}$, $E_{\rm K,iso}=(2.28\pm0.27)\times10^{53}$ erg, $t_s\sim 2400$ s, $t_e=2997\pm 546$ s, $L_0=(4.0\pm2.5)\times10^{50}$ erg/s, $q=-0.18_{-0.07}^{+0.05}$, and $\theta_{\rm j}>0.15$ rad. The flare-like X-ray data at around $10^3$ are not included in our fits.}
\label{model_injection}
\end{figure}

\begin{itemize}
\item Constraining the medium property and the power-law index of the radiating electrons with the closure relation of the forward shock model. With the decay slope and spectral index of the normal decay phase (Phase V), we find that the afterglows are radiated in the spectral regime of $\nu_{m}<\nu<\nu_{c}$. In this regime we have $p=2\beta+1$, where $\beta=\Gamma-1$. We therefore obtain $p=2.44\pm 0.06$. We fix $p=2.4$ in our analysis without considering the uncertainty of $p$. Note that the slope of the afterglow onset (Phase I) is $\alpha_1=2.6\pm 0.4$, which well agrees with the expectation for a constant density interstellar medium (ISM). The medium density in our fit then is set as a constant $n$.
\item Describing the energy injection as $L_{\rm in}=L_0 t^{q}$ during a period from the starting ($t_s$) to the ending ($t_e$) time in order to explain the Phase IV.
\item Adopting the Markov Chain Monte Carlo (MCMC) technique to search for the parameter set that can best represent the data. The parameters of our model include the initial Lorentz factor ($\Gamma_0$), the fraction of shock energy to electron ($\epsilon_e$), the fraction of shock energy to magnetic field($\epsilon_B$), the medium density ($n$), the isotropic kinetic energy ($E_{\rm K, iso}$), the jet opening angle ($\theta_j$), and the parameters of the energy injection ($L_0$, $q$, $t_s$, and $t_e$). They are set in the following ranges, $\Gamma_0\in[50,150]$,$\epsilon_e\in [0.01,0.5]$, $\epsilon_B\in [10^{-7},10^{-4}]$\footnote{Some recent statistical analysis suggests a low $\epsilon_B$, i.e., $[10^{-6},10^{-3}]$ (e.g., Wang et al. 2015; Gao et al. 2015; Japelj et al. 2014). Therefore, we set $\epsilon_B\in [10^{-7}, 10^{-2}]$. We find that a reasonable parameter set that can roughly represent the optical and XRT lightcurves requires $\epsilon_B<10^{-4}$. We then finalize our fit by setting $\epsilon_B\in [10^{-7},10^{-4}]$.}, $n\in [0.1,25]$, $E_{\rm K,iso}\in[10^{51},10^{54}]$ erg, $\theta_{\rm j}\in [0.01,0.5]$ rad, $t_s\in[1000,3000]$ seconds, $t_e\in [3000,5000]$ seconds, $L_0\in [10^{49}, 10^{52}]$ erg/s, and $q\in [-0.3,-0.1]$. We calculate the $\chi^2$ and measure the goodness of the fits for each parameter set with a normalized probability $p_f\propto e^{-\chi^2/2}$. Note that the lightcurves are composed of some flares. With the MCMC technique£¬ we search for the parameter set that have the minimum $\chi^2$ (hence the largest $p_f$ value). The uncertainty of a parameter in the best parameter set is evaluated by fixing the other parameters.
\end{itemize}
With this strategy, the best parameters and their uncertainty (1$\sigma$ confidence level) are $\Gamma_0=91\pm8$, $\epsilon_e=0.086\pm0.008$, $\epsilon_B=(1.64\pm0.25)\times10^{-6}$, $n=13.3\pm2.6$  cm$^{-3}$, $E_{\rm K,iso}=(2.28\pm0.27)\times10^{53}$ erg, $t_s\sim 2400$ s, $t_e=2997\pm 546$ s, $L_0=(4.0\pm2.5)\times10^{50}$ erg/s, and $q=-0.18_{-0.07}^{+0.05}$. The jet opening angle is poorly constrained and we have $\theta_{\rm j}>0.15$ rad. Figure \ref{model_injection} shows our best fit to the data with our model. The $\chi^2$ of the fit is 1.605. The large $\chi^2$ is due to flares/fluctuations in the optical and X-ray bands. The derived $\epsilon_e$ is generally consistent with previous results (Wijers \&
Galama 1999; Panaitescu \& Kumar 2001; Yost et al. 2003; Liang
et al. 2004), but $\epsilon_B$ is much lower than the typical value, i.e., $10^{-2}$ (e.g.,
Panaitescu \& Kumar 2001). Further discussion on $\epsilon_B$ is presented in \S 5.1. The $\Gamma_0$ of GRB 110530A is at the lower end of the $\Gamma_0$ distribution for a sample of GRBs whose $\Gamma_0$ values are calculated with the deceleration time in their optical afterglow lightcurves (see Figure 12 of Liang et al. 2013).

{Note that the redshift of GRB 110530A is unknown, we set $z=1$ in our lightcurve modeling\footnote{Liang et al. (2015) found a tight correlation among $L_{\rm \gamma, iso}$,  $\Gamma_0$, and $E_{p}$ in the burst frame, i.e., $\log L_{\rm \gamma, iso}/10^{52} {\rm erg \ s^{-1}}=(-6.38\pm 0.35)+(1.34\pm 0.14)\times \log (E_{\rm p}(1+z))+(1.32\pm 0.19)\times \log \Gamma_0$. By setting $z=1$ and using $E_{\rm p}=45$ keV and $\Gamma_0=91$, we get $\log L_{\rm \gamma, iso}/{\rm erg\ s^{-1}}=50.82\pm 0.35$ in the energy band of $1-10^4$ keV, where the error is measured only for the systematical error of the relation without considering the observed errors of $E_p$ and $\Gamma_0$ since no $E_p$ error is available. This is roughly consistent with the observed luminosity by correcting to the same energy band, i.e., $\log L^{\rm c}_{\rm iso, obs}/{\rm erg\ s^{-1}}=50.45\pm 0.11$}. We also check the dependence of the model parameters on the burst distance by setting $z=0.5$ and $z=2.0$. It is found that $\epsilon_e$, $\epsilon_B$, $n$, $q$ do not change with redshift. Being due to large uncertainties of $t_e$,$t_s$ and $\theta_j$, we also do not find clear dependence of these parameters on redshift. However, $\Gamma_0$, $E_{\rm K, iso}$, and $L_0$  are getting larger as $z$ increases.}

\section{Discussion}

\subsection{Baryonic or Magnetized Jet?}
The issue that GRB jets are baryonic or magnetized is under debating (e.g., Zhang 2011). The GRB radiative efficiency, which is defined as $\eta_\gamma=E_{\gamma,iso}/(E_{\gamma,\rm iso}+E_{K,\rm iso})$, is an essential quantity to understand the nature of the bursts (e.g., Zhang et al. 2006). The standard internal shock models predict a GRB efficiency
of $\sim 1\%$ (Kumar 1999; Panaitescu et al. 1999). $E_{\rm K}$ should be the kinetic energy of the fireball that produces the observed gamma-ray energy and it would be estimated at the fireball deceleration time. Assuming that the early optical bump is due to the fireball deceleration by the ambient medium, one then can derive the $E_{\rm K}$ of the fireball at the deceleration time ($t_{\rm dec}$) by eliminating the possible late energy injection. In this analysis, we get $E_{\rm K,iso}=(2.28\pm0.27)\times10^{53}$ erg. The corrected gamma-ray energy in $1-10^4$ keV band is $E^{\rm c}_{\rm \gamma, iso}=1.92\times 10^{51}$ erg. Therefore, the internal shock radiation efficiency of GRB 110530A is $0.83\%$. The total energy injection from $2390/(1+z)$ to $2997/(1+z)$ seconds derived from our model fit is $\sim 3.39\times 10^{52}$ erg. Including the delayed energy injection, the efficiency is $\eta=0.73\%$. This is also consistent with the prediction of the internal shock models. Zhang et al. (2007a) found that some bursts have a low efficiency throughout, and these GRBs usually have an X-ray afterglow light curve that smoothly joins the prompt emission light curve without a distinct steep decay component or an extended shallow decay component. Fan \& Piran (2006) suggested that the gamma-ray efficiency is moderate and does not challenge the standard internal shock model. GRB 110530A is consistent with that reported by Zhang et al. (2007a). The low efficiency well agrees with the prediction of the standard internal shock models, likely implying that the outflow for the prompt emission could be baryonic.

The jet composition in the afterglow phase is also of interest. The $\epsilon_B$ value derived from our model fit is much smaller than the typical value reported in the literature. For a constant density medium, the cooling frequency of synchrotron emission
frequency is given by $\nu_c =6.3 \times 10^{15} {\rm\ Hz}
(1+z)^{-1/2} (1+Y)^{-2} \epsilon_{B,-2}^{-3/2}E_{\rm K,iso, 52}^{-1/2} n^{-1}
t_d^{-1/2}$ (Sari et al. 1998; Yost et al. 2003), where $Y$ is the Inverse Compton scattering parameter and $t_d$ is the observer's time in unit of days. One can see $\nu_c$ is sensitive to $\epsilon_B$. As time increases, $\nu_c$ is getting smaller. The extremely low $\epsilon_B$ ensures that both the optical and X-ray emission is still in the regime $\nu<\nu_c$ for the the time at several days. The magnetic field strength of the afterglow jet in the co-moving frame is given by $B=(32\pi m_p\epsilon_B n)^{1/2}\Gamma_0 c$, and the power carried by the magnetic field can be derived from $P_B=\pi R_{\rm dec} c B^{2}/8\pi$, where $R_{\rm dec}=2.25\times 10^{16} (\Gamma_{0}/100)^2 (t_{p,z}/100 {\rm s})$ is the deceleration radius of the fireball, $m_p$ is the proton mass, and $c$ is the speed of light. We obtain $B=0.165$ G and $P_B\sim 5.55\times 10^{44}$ erg/s for GRB 110530A. Assuming that the electron energy is full radiated and the X-ray luminosity is a good representative of the bolometric afterglow luminosity, we estimate the kinetic power of the afterglow jet at the deceleration time with $L_{\rm K}=L_{\rm X}(1-\cos\theta_j)/\epsilon_e$, which gives $L_{\rm K} > 1.33\times 10^{46}$ erg/s. Therefore, the magnetization of the afterglow jet is $\sigma=P_B/L_{\rm K}< 0.04$, suggesting that the afterglow jet is baryonic.
It is also interesting that the derived $B$ and $\sigma$ are comparable to the typical values of the jets in BL Lacs, which are suggested to be matter dominated (Zhang et al. 2013)

\subsection{Possible Sources of the Delayed Energy Injection}
 A plateau phase is usually observed in the XRT lightcurves (Zhang et al. 2006; O'Brien et al. 2006; Liang et al. 2007) and in about one-third of optical lightcurves for long-duration GRBs (Li et al. 2012; Liang et al. 2013). Such a feature can be well explained with the long-lasting energy injection from a constant magnetic-dipole-radiation luminosity within the spin-down timescale of a magnetar (Dai \& Lu 1998; Zhang \& M\'{e}sz\'{a}ros 2001; L\"{u} \& Zhang 2014). The injection behavior in this scenario is continuous and starts at a very early epoch. With clear detection of the afterglow onset bump, we propose that the energy injection could be happened post the deceleration time of the fireball. In addition, as shown above, the jet in the prompt gamma-ray phase and afterglow phase seem to be matter dominated. These results possibly disfavor the scenario pulsar wind injection\footnote{It was also proposed that the magnetic-dipole-radiation luminosity of a magnetar can dramatically increase with time, which may lead to a significant bump in the afterglow lightcurves, if the magnetar is spun up by the accretion matter (Dai \& Liu 2012). In this scenario, the energy injection in early epoch is not significant and may feature as delayed energy injection in late epoch.}.
 We suggest that the injection may caused by a slower shallow that is ejected at the same epoch as that of the shells for producing the prompt gamma-rays (Zhang \& M\'{e}sz\'{a}ros 2002) or delayed ejecta from late accretion activity (Geng et al. 2013). The time delay of the rear shells/ejecta for catching up with the decelerated fireball may result in the delayed energy injection. On the other hand, the energy transfer time from fireball ejecta to ambient medium typically extends to thousands of seconds, which may also broaden the onset peak in the thin shell case (Kobayashi \& Zhang 2007).

In the scenario of a black hole accretion system, the energy flow from the fall-back accretion may be delayed for a
fall-back time $t_{\rm fb}$ and produce giant bumps in the optical bands (Geng et al. 2013). In this scenario, one may place some constraint on the progenitor stars. The radius of the fall-back material can be estimated with $R_{\rm fb}\sim 6.85\times 10^{10} {\rm cm} (M_{\rm BH}/3M_\odot)^{1/3} (t_{\rm fb}/10^3{\rm s})^{2/3}$. We estimate the minimum and maximum radii of the fall-back material with the $t_{s}$ and $t_e$ in the burst frame and have $R_{\rm fb, \min}\sim 7.71\times 10^{10} {\rm cm} (M_{\rm BH}/3M_\odot)^{1/3}$ and $R_{\rm fb, \max}\sim 8.98 \times 10^{10} {\rm cm} (M_{\rm BH}/3M_\odot)^{1/3}$. Woosley \& Weaver (1995) derived the mass density profile as a function
of radius $R$ with simulations for a pre-supernova star with mass of 25M$_\odot$ (see also Janiuk \& Proga 2008), as shown in Figure \ref{mass_profile}. The mass density of the shell $R\in [R_{\rm fb, \min}, R_{\rm fb, \max}]$ is about $1.7\times 10^{-2}$ g cm$^{-3}$, and the mass in this shell is $9.62\times 10^{-3}$ $M_{\odot}$ (corresponding to an energy of $1.71\times 10^{52}$ erg), assuming that $M_{\rm BH}=3M_{\odot}$. The total energy injection from 2390 seconds to 2997 seconds derived from our model fit is $\sim 3.39\times 10^{52}$ erg, corresponding to a geometrically-corrected injection energy of $3.81\times 10^{50}$ erg by taking $\theta_j=0.15$ rad. The jet radiation is only a small fraction ($2.23\%$) of the fall-back mass. {By simplifying the mass density profile as a power-law function, $\log \rho/{\rm g\ cm^{-3}}=30.47-3.24\log R/{\rm cm}$ within $R<R_{\rm fb, \max}$, as shown in Figure 5, the mass within $R<R_{\rm fb, \max}$ is $\sim 7.5M_\odot$. If all the mass within $R<R_{\rm fb, \max}$ is collapsed to form a newly-born black hole and its accretion disk, the total collapsed/fall-back mass is about a fraction of 30\% of the progenitor star, and the rest mass in other outer layers would be broken out as a supernova.}

\section{Conclusions}
We have reported our very early optical observations for GRB 110530A and investigate its jet properties together with its X-ray afterglow data. A broad bump with significant flares is observed in the optical lightcurve at $t<2000$ seconds, which is followed by a plateau with transition to a normal decaying segment at $t=3000$ seconds. The X-ray afterglow lightcurve shows almost the same feature. Our joint spectral fits of the optical and X-ray data show that they are in the same regime, with a photon index of $\sim 1.70$. The extinction of the host galaxy is negligible, but the equivalent hydrogen column density of host galaxy is approximately $1.0\times 10^{21}$cm$^{-2}$. We model the optical and X-ray lightcurves with the standard external shock model by considering delayed energy injection and assuming its redshift as 1. Our best parameters derived from a MCMC approach are $\Gamma_0=91\pm 8$, $\epsilon_e=0.086\pm 0.008$, $\epsilon_B=(1.64\pm 0.25)\times 10^{-6}$, $n=13.3\pm 2.6$ cm$^{-3}$, $E_{\rm K, iso}=(2.28\pm 0.27)\times 10^{53}$ erg, and $\theta_j\sim 0.15$ rad. The energy injection can be described as $L_{\rm in}/10^{50}{\rm erg\ s^{-1}}=(4.0\pm 2.5)\times t^{-0.18}$, which starts at $\sim 2390$ seconds and lasts only about 700 seconds. Based on our modeling results, the radiative efficiency of the GRB fireball is $\sim 1\%$, the magnetic field strength and the magnetization parameter of the afterglow jet are $B=0.165$ G and $\sigma<0.04$, respectively. We propose that the jet would be matter dominated and possible sources of the delayed energy injection are also discussed.

The most striking observation of GRB 110530A is its early broad bump following by a plateau in its R band afterglow lightcurve. We have shown that the standard forward shock model with a delayed injection can roughly fit the global feature of the lightcurves. We address the flickerings in the optical and X-ray lightcurves as superimposed flares that may have internal origins. We should note that these flickering, especially the significant flickering at around 3000 seconds in the X-ray band, may be also due to the delayed energy injection. Zhang \& M\'{e}sz\'{a}ros (2001) analyzed the energy injection and corresponding signature that could be shown up in afterglow lightcurves. They showed that injection by a Poynting-flux-dominated shell that has an energy comparable to that of the initial fireball would lead to a gradual achromatic bump. In the case when the injection is kinetic-energy-dominated, the results depend on the situation of the collision between the injected (rear) shells and initial (leading) shells. If the collision is mild, the signature showed in the lightcurves may be analogous to the Poynting-flux-dominated injection case. In case of a violent collision a significant flare-like bump may be observed (see Figure 5 of Zhang \& M\'{e}sz\'{a}ros 2001). In the case that the delayed energy injection is fed by the fall-back materials, the delayed energy would also cause a notable
rise to the Lorentz factor of the external shock, which will ¡°generate¡± a bump in the multiple band afterglows as seen in  GRB 081029 and GRB 100621A (Nardini et al. 2011; Greiner et al. 2013; Geng et al. 2013).

\section{Acknowledgement}
This work is supported by the National Basic Research Program of China (973 Program, grant No. 2014CB845800),
the National Natural Science Foundation of China (Grant No. 11533003, 11103036, U1331202, U1231115 and U1331101), the Strategic Priority Research Program ¡°The Emergence of Cosmological Structures¡± of the Chinese Academy of Sciences (grant XDB09000000), the Guangxi Science Foundation (Grant No. 2013GXNSFFA019001). We also acknowledge the use of the public data from the Swift data archive.

\begin{deluxetable}{cccccc}
\tabletypesize{\small}
\tablewidth{0pt}
\label{Tab:data}
\tablecaption{Optical Afterglow Photometry Log of GRB 110530A}
\tablehead{
\colhead{T-T0(mid,second)} &
\colhead{Exposure (sec)} &
\colhead{Mag$^{a}$} &
\colhead{$\sigma^{a}$} &
\colhead{Filter} &
\colhead{Telescope}
}
\startdata
\object{	144	}&	20	&	19.24	&	0.34	&	W	&	TNT	\\
\object{	167	}&	20	&	19.39	&	0.24	&	W	&	TNT	\\
\object{	190	}&	20	&	18.99	&	0.22	&	W	&	TNT	\\
\object{	213	}&	20	&	19.01	&	0.12	&	W	&	TNT	\\
\object{	235	}&	20	&	18.86	&	0.18	&	W	&	TNT	\\
\object{	258	}&	20	&	18.59	&	0.14	&	W	&	TNT	\\
\object{	281	}&	20	&	18.63	&	0.14	&	W	&	TNT	\\
\object{	303	}&	20	&	18.54	&	0.14	&	W	&	TNT	\\
\object{	326	}&	20	&	18.56	&	0.14	&	W	&	TNT	\\
\object{	349	}&	20	&	18.54	&	0.10	&	W	&	TNT	\\
\object{	372	}&	20	&	18.60	&	0.12	&	W	&	TNT	\\
\object{	394	}&	20	&	18.54	&	0.15	&	W	&	TNT	\\
\object{	417	}&	20	&	18.66	&	0.16	&	W	&	TNT	\\
\object{	440	}&	20	&	18.53	&	0.13	&	W	&	TNT	\\
\object{	463	}&	20	&	18.27	&	0.12	&	W	&	TNT	\\
\object{	485	}&	20	&	18.41	&	0.11	&	W	&	TNT	\\
\object{	508	}&	20	&	18.43	&	0.14	&	W	&	TNT	\\
\object{	531	}&	20	&	18.36	&	0.13	&	W	&	TNT	\\
\object{	553	}&	20	&	18.48	&	0.12	&	W	&	TNT	\\
\object{	605	}&	60	&	18.62	&	0.09	&	R	&	TNT	\\
\object{	684	}&	60	&	18.47	&	0.08	&	R	&	TNT	\\
\object{	763	}&	60	&	18.45	&	0.07	&	R	&	TNT	\\
\object{	841	}&	60	&	18.49	&	0.10	&	R	&	TNT	\\
\object{	919	}&	60	&	18.37	&	0.08	&	R	&	TNT	\\
\object{	998	}&	60	&	18.60	&	0.10	&	R	&	TNT	\\
\object{	1076	}&	60	&	18.64	&	0.12	&	R	&	TNT	\\
\object{	1155	}&	60	&	18.49	&	0.08	&	R	&	TNT	\\
\object{	1233	}&	60	&	18.62	&	0.11	&	R	&	TNT	\\
\object{	1312	}&	60	&	18.67	&	0.12	&	R	&	TNT	\\
\object{	1390	}&	60	&	18.76	&	0.13	&	R	&	TNT	\\
\object{	1469	}&	60	&	18.83	&	0.11	&	R	&	TNT	\\
\object{	1547	}&	60	&	19.17	&	0.18	&	R	&	TNT	\\
\object{	1625	}&	60	&	18.92	&	0.14	&	R	&	TNT	\\
\object{	1704	}&	60	&	19.08	&	0.14	&	R	&	TNT	\\
\object{	1782	}&	60	&	19.01	&	0.16	&	R	&	TNT	\\
\object{	1861	}&	60	&	19.03	&	0.14	&	R	&	TNT	\\
\object{	1939	}&	60	&	19.27	&	0.23	&	R	&	TNT	\\
\object{	2018	}&	60	&	19.17	&	0.17	&	R	&	TNT	\\
\object{	2096	}&	60	&	19.53	&	0.24	&	R	&	TNT	\\
\object{	2297	}&	300	&	19.33	&	0.08	&	R	&	TNT	\\
\object{	2614	}&	300	&	19.46	&	0.08	&	R	&	TNT	\\
\object{	2932	}&	300	&	19.40	&	0.07	&	R	&	TNT	\\
\object{	3250	}&	300	&	19.44	&	0.08	&	R	&	TNT	\\
\object{	3567	}&	300	&	19.36	&	0.07	&	R	&	TNT	\\
\object{	3885	}&	300	&	19.44	&	0.09	&	R	&	TNT	\\
\object{	4203	}&	300	&	19.53	&	0.09	&	R	&	TNT	\\
\object{	4520	}&	300	&	19.65	&	0.09	&	R	&	TNT	\\
\object{	4838	}&	300	&	19.50	&	0.08	&	R	&	TNT	\\
\object{	5156	}&	300	&	19.84	&	0.10	&	R	&	TNT	\\
\object{	5473	}&	300	&	19.70	&	0.10	&	R	&	TNT	\\
\object{	6109	}&	300	&	19.68	&	0.09	&	R	&	TNT	\\
\object{	6427	}&	300	&	19.82	&	0.10	&	R	&	TNT	\\
\object{	6744	}&	300	&	19.73	&	0.10	&	R	&	TNT	\\
\object{	7062	}&	300	&	20.04	&	0.11	&	R	&	TNT	\\
\object{	7380	}&	300	&	20.13	&	0.12	&	R	&	TNT	\\
\object{	7697	}&	300	&	20.27	&	0.14	&	R	&	TNT	\\
\object{	8015	}&	300	&	20.26	&	0.14	&	R	&	TNT	\\
\object{	8333	}&	300	&	20.19	&	0.13	&	R	&	TNT	\\
\object{	8650	}&	300	&	20.23	&	0.15	&	R	&	TNT	\\
\object{	8968	}&	300	&	20.34	&	0.15	&	R	&	TNT	\\
\object{	9286	}&	300	&	20.09	&	0.12	&	R	&	TNT	\\
\object{	9604	}&	300	&	20.11	&	0.12	&	R	&	TNT	\\
\object{	10371	}&	600	&	20.49	&	0.10	&	R	&	TNT	\\
\object{	10557	}&	300	&	20.30	&	0.14	&	R	&	TNT	\\
\object{	11624	}&	900	&	20.51	&	0.10	&	R	&	TNT	\\
\object{	13177	}&	1500	&	20.66	&	0.09	&	R	&	TNT	\\
\object{	14166	}&	900	&	21.00	&	0.20	&	R	&	TNT	\\
\object{	15719	}&	1500	&	21.17	&	0.46	&	R	&	TNT	\\
\object{	79442	}&	3000	&	$>$21.91	&	---	&	R	&	TNT	\\
\object{	1835	}&	300	&	19.59	&	0.14	&	g	&	LOT	\\
\object{	2814	}&	300	&	19.82	&	0.17	&	g	&	LOT	\\
\object{	3792	}&	300	&	19.78	&	0.24	&	g	&	LOT	\\
\object{	4771	}&	300	&	19.92	&	0.15	&	g	&	LOT	\\
\object{	5749	}&	300	&	20.11	&	0.17	&	g	&	LOT	\\
\object{	6727	}&	300	&	20.31	&	0.16	&	g	&	LOT	\\
\object{	2486	}&	300	&	19.11	&	0.15	&	i	&	LOT	\\
\object{	3465	}&	300	&	19.05	&	0.19	&	i	&	LOT	\\
\object{	4443	}&	300	&	19.14	&	0.17	&	i	&	LOT	\\
\object{	5421	}&	300	&	19.26	&	0.17	&	i	&	LOT	\\
\object{	6399	}&	300	&	19.24	&	0.21	&	i	&	LOT	\\
\object{	7377	}&	300	&	19.25	&	0.18	&	i	&	LOT	\\
\object{	1381	}&	300	&	19.02	&	0.12	&	r	&	LOT	\\
\object{	2161	}&	300	&	19.59	&	0.14	&	r	&	LOT	\\
\object{	3139	}&	300	&	19.52	&	0.14	&	r	&	LOT	\\
\object{	4118	}&	300	&	19.75	&	0.21	&	r	&	LOT	\\
\object{	5096	}&	300	&	19.87	&	0.15	&	r	&	LOT	\\
\object{	6074	}&	300	&	19.91	&	0.14	&	r	&	LOT	\\
\object{	7052	}&	300	&	20.18	&	0.16	&	r	&	LOT	\\
\object{	512	}&	74	&	20.30	&	0.14	&	white	&	UVOT	\\
\object{	939	}&	74	&	20.32	&	0.14	&	white	&	UVOT	\\
\object{	7748	}&	546	&	21.48	&	0.20	&	white	&	UVOT	\\
\object{	13369	}&	707	&	22.07	&	0.25	&	white	&	UVOT	\\
\object{	17892	}&	1976	&	$>$22.88	&	---	&	white	&	UVOT	\\
\object{	21569	}&	1230	&	$>$22.83	&	---	&	white	&	UVOT	\\
\enddata
\tablecomments{(a)Not corrected for Galactic foreground reddening.\\
The reference time $T_0$ is {\em Swift} BAT burst trigger time.\\
"T-T0" is the middle time in second for each data.\\
"Exposure" is the exposure time for each data in second.\\
"$\sigma$" means the uncertainty of the magnitude.\\}

\end{deluxetable}

\begin{deluxetable}{cccccc}
\tabletypesize{\small}
\tablewidth{300pt}
\label{Tab:ReferenceStars}
\tablecaption{Reference stars for magnitude calibration}
\tablehead{
\colhead{RA} &
\colhead{DEC} &
\colhead{Epoch} &
\colhead{$B2$} &
\colhead{$R2$} &
\colhead{$I$}
}
\startdata
18:48:17.785 &  +61:55:56.69 &  J2000 &   18.63 &  17.07 &  16.08 \\
18:48:15.583 &  +61:56:13.39 &  J2000 &   17.48 &  16.08 &  14.86 \\
18:48:15.951 &  +61:56:25.06 &  J2000 &   18.41 &  17.23 &  17.02 \\
18:48:10.257 &  +61:55:43.25 &  J2000 &   17.49 &  16.05 &  15.08 \\
18:48:08.206 &  +61:55:40.41 &  J2000 &   17.09 &  16.88 &  16.30 \\
18:48:05.743 &  +61:54:51.29 &  J2000 &   17.41 &  16.93 &  16.37 \\
18:48:19.011 &  +61:54:43.98 &  J2000 &   15.15 &  14.03 &  13.20 \\
18:48:23.664 &  +61:55:10.04 &  J2000 &   16.11 &  15.34 &  14.80 \\
18:48:27.258 &  +61:55:12.79 &  J2000 &   16.17 &  15.36 &  14.30 \\
18:48:26.344 &  +61:56:20.09 &  J2000 &   16.24 &  15.77 &  15.58 \\
18:48:22.675 &  +61:56:37.35 &  J2000 &   16.49 &  16.10 &  15.47 \\
\enddata
\tablecomments{
Reference stars for the calibration in this work. $B2$,
$R2$ and $I$-band magnitudes are extracted
from USNO B1.0 catalog.}
\end{deluxetable}

\begin{deluxetable}{cccc}
\tabletypesize{\footnotesize}
\tablewidth{400pt}
\label{Tab:xrt-opt}
\tablecaption{Spectral analysis of the Optical and X-ray Afterglows in selected time intervals}
\tablehead{
\colhead{Interval(s)} &
\colhead{Model($\chi^{2}/{\rm dof}$)} &
\colhead{PhoIndex($\Gamma$)} &
}
\startdata
\object{0.6k-0.9k}  &  LMC*PL($7.54/7=1.08$)     &    $1.70\pm0.02$     \\
                    &  SMC*PL($7.54/7=1.08$)     &    $1.70\pm0.02$     \\
\object{0.9k-1.37k} &  LMC*PL($15.79/13=1.214$)  &    $1.68\pm0.13$     \\
                    &  SMC*PL($15.79/13=1.215$)  &    $1.68\pm0.13$     \\
\object{1.37k-2.5k} &  LMC*PL($33.83/22=1.538$)  &    $1.72\pm0.04$     \\
                    &  SMC*PL($33.83/22=1.538$)  &    $1.72\pm0.04$     \\
\object{6k-9k}      &  LMC*PL($31.58/11=2.871$)  &    $1.67\pm0.02$     \\
                    &  SMC*PL($31.59/11=2.872$)  &    $1.67\pm0.02$     \\
\object{9k-14k}     &  LMC*PL($3.03/4=0.758$)    &    $1.72\pm0.03$     \\
                    &  SMC*PL($3.03/4=0.758$)    &    $1.72\pm0.03$     \\

\enddata
\end{deluxetable}
\clearpage

\begin{figure}
\includegraphics[angle=0,scale=0.2]{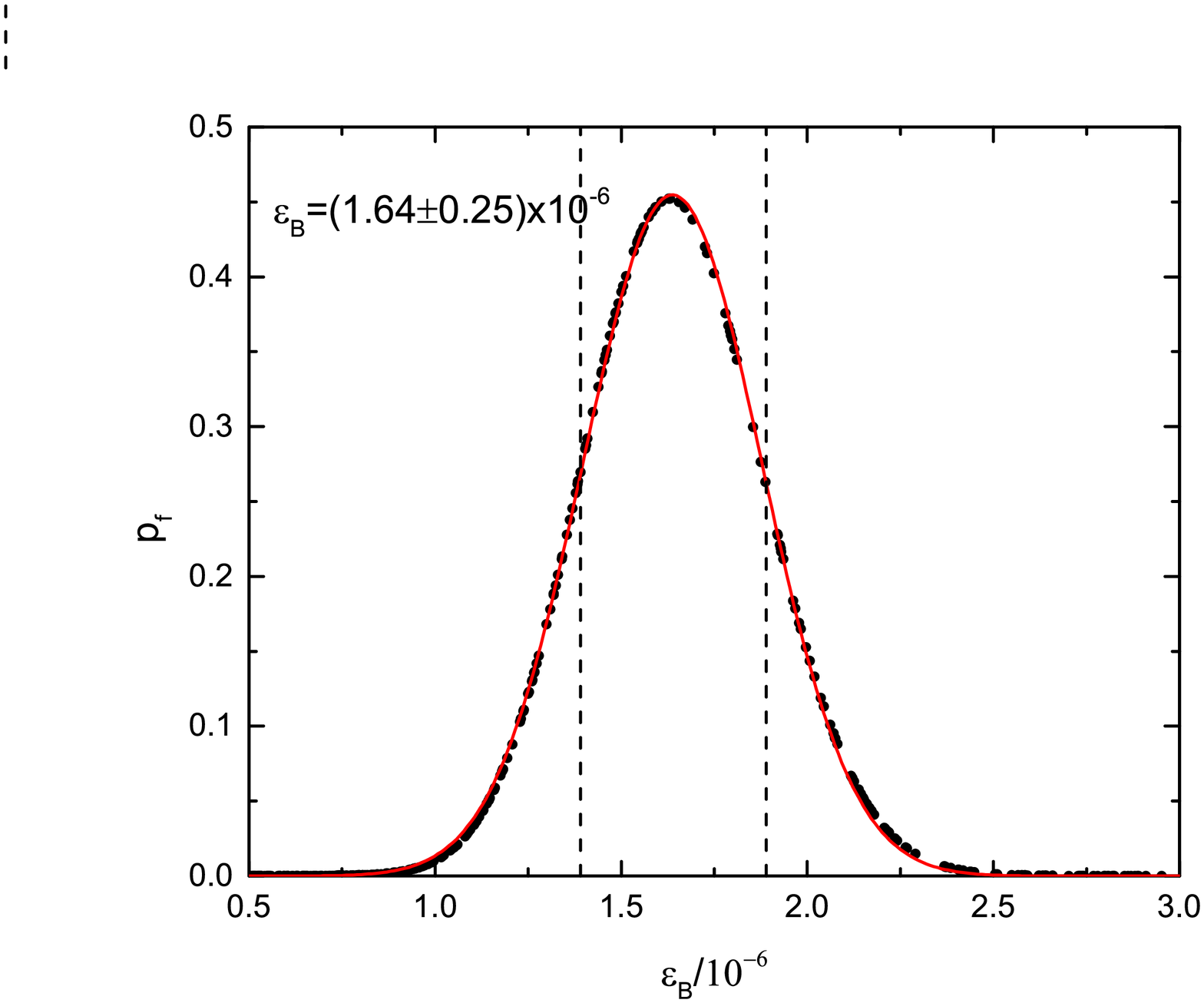}
\includegraphics[angle=0,scale=0.2]{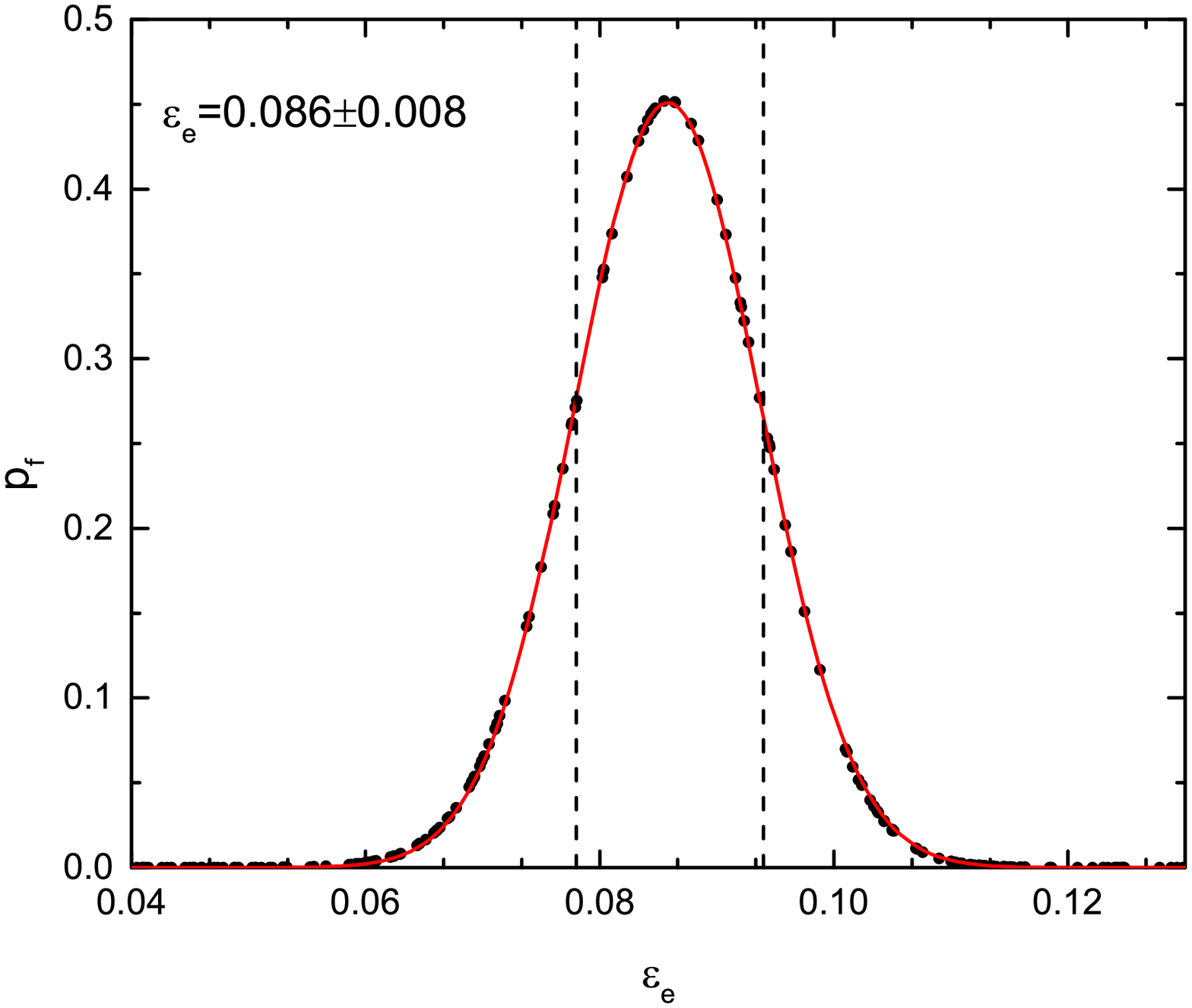}
\includegraphics[angle=0,scale=0.2]{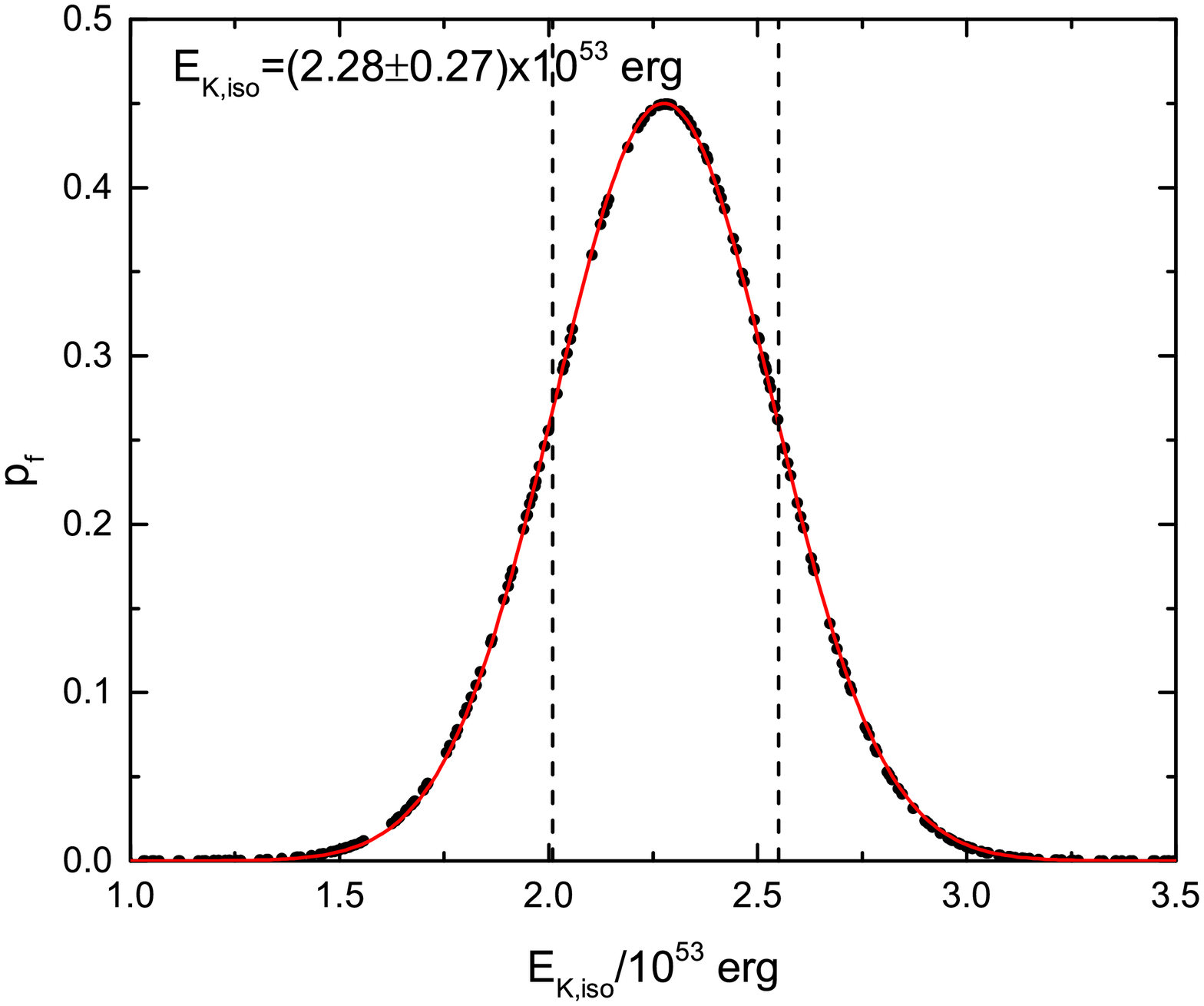}
\includegraphics[angle=0,scale=0.2]{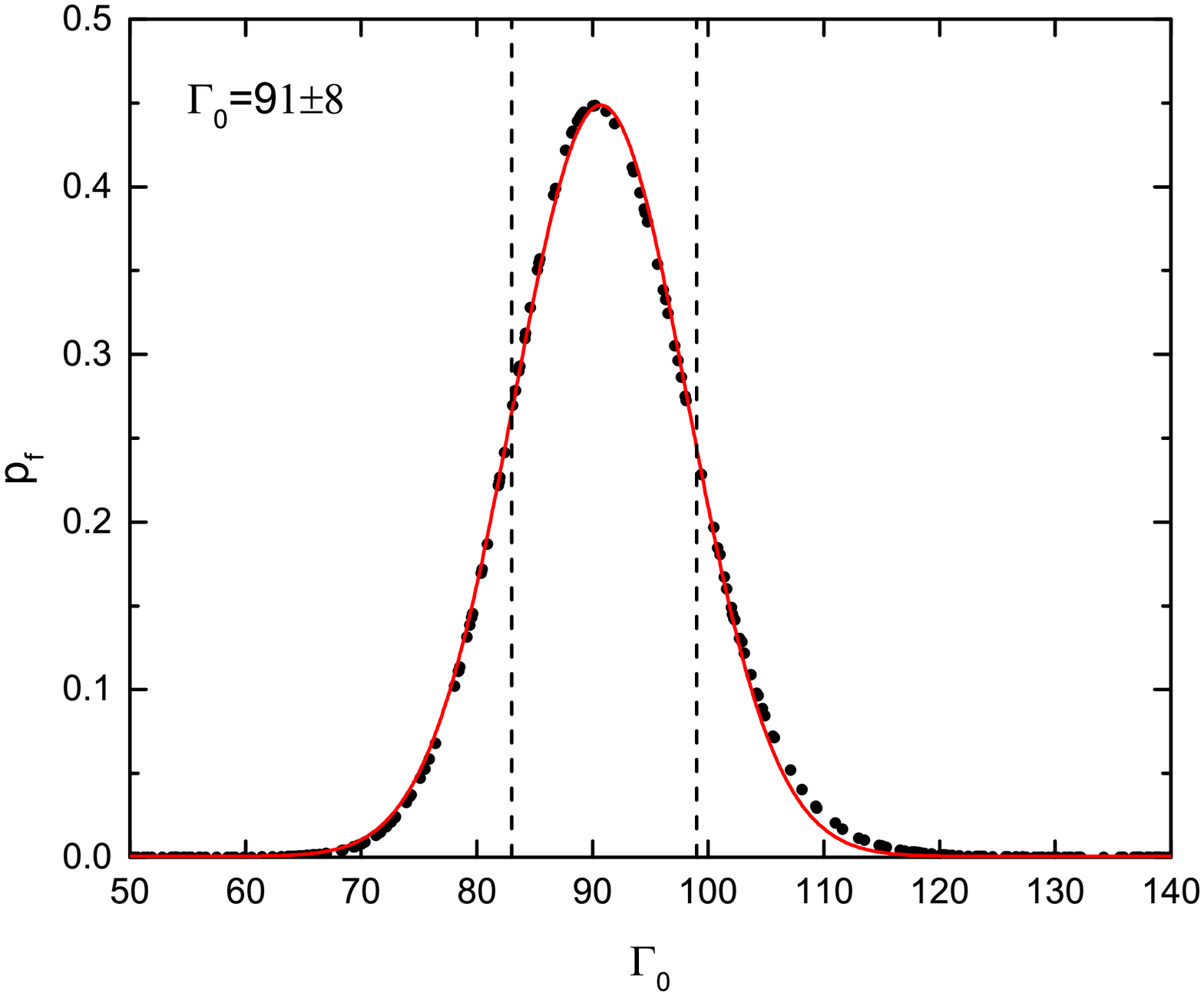}
\includegraphics[angle=0,scale=0.2]{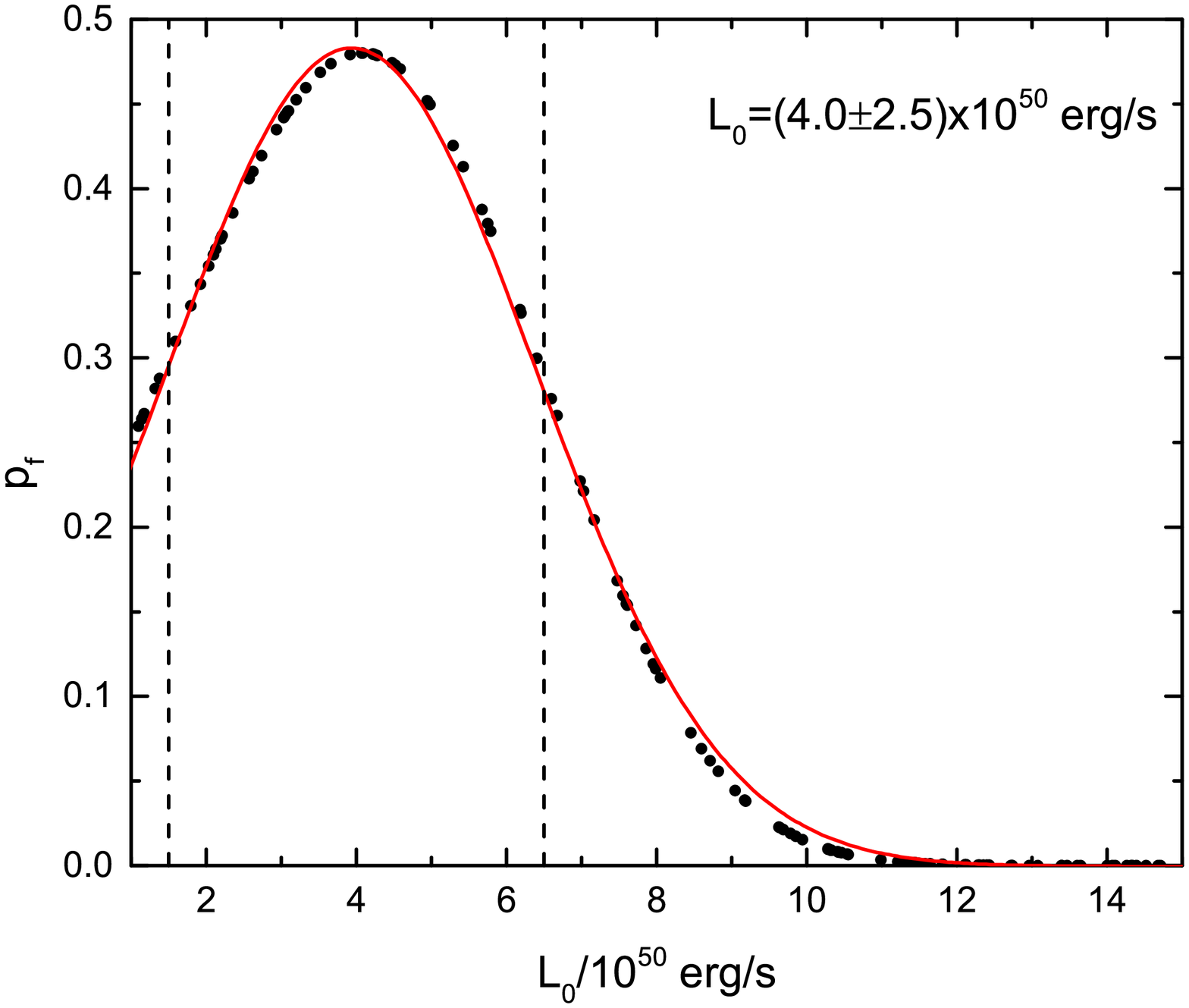}
\includegraphics[angle=0,scale=0.2]{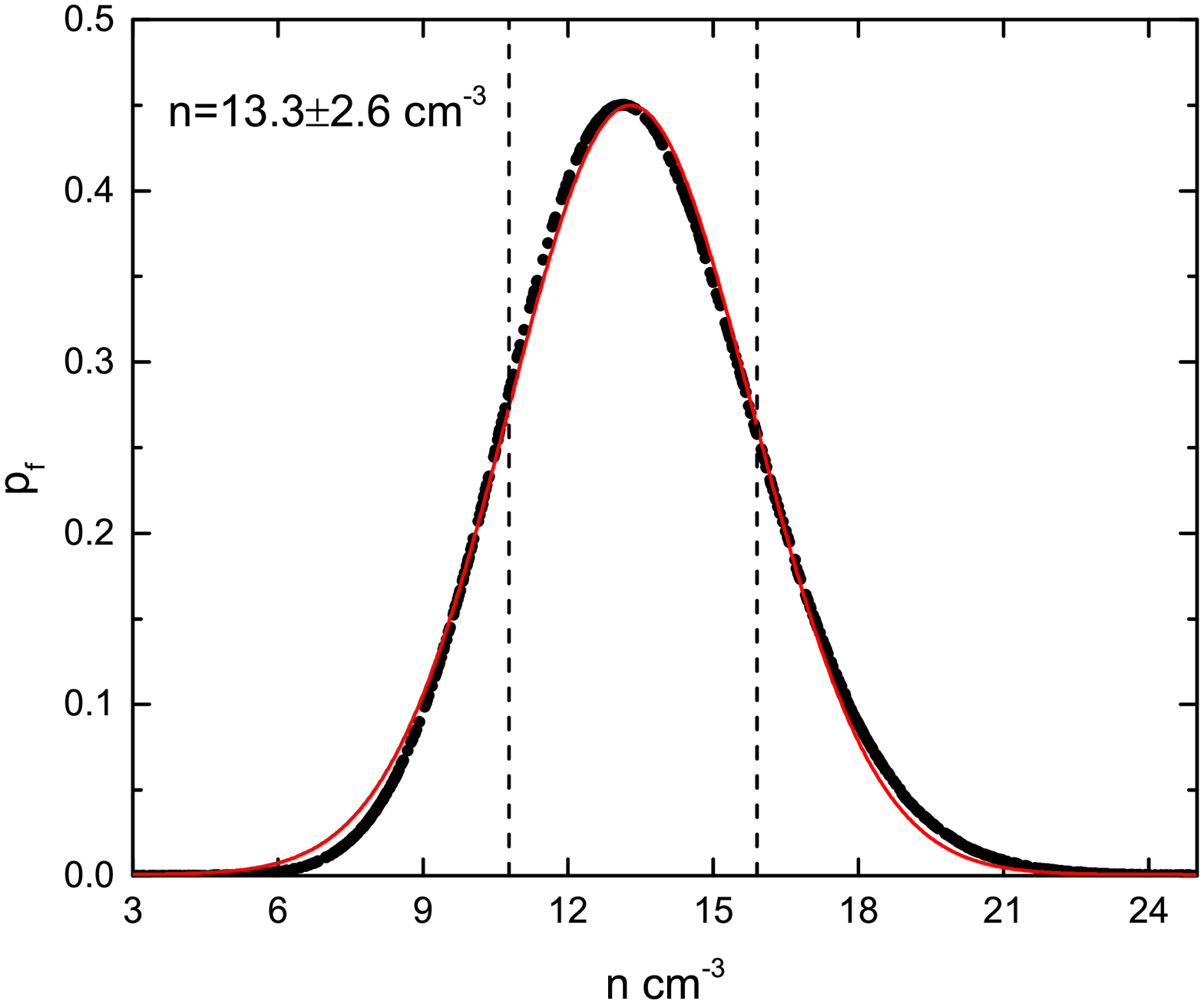}
\includegraphics[angle=0,scale=0.2]{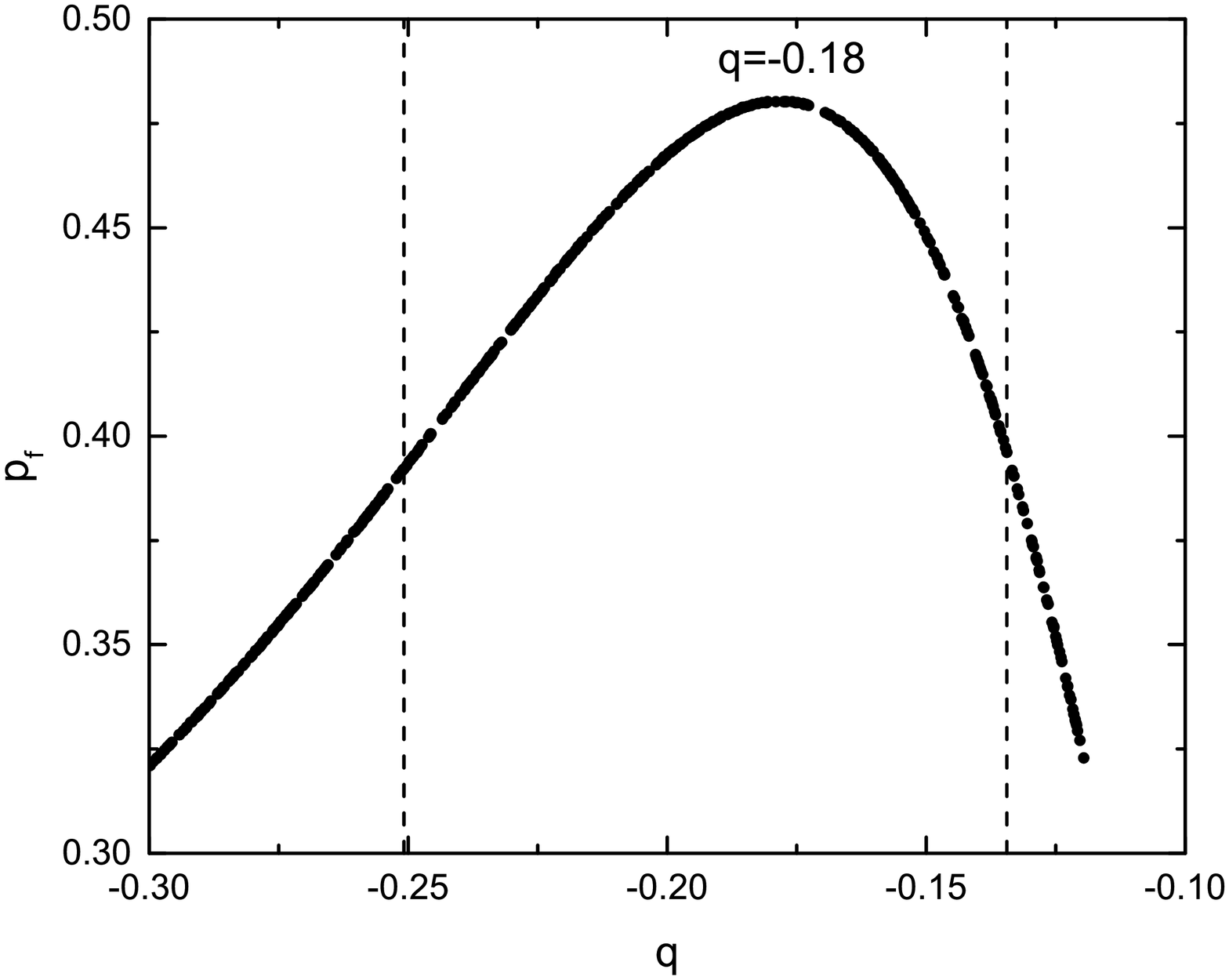}
\includegraphics[angle=0,scale=0.2]{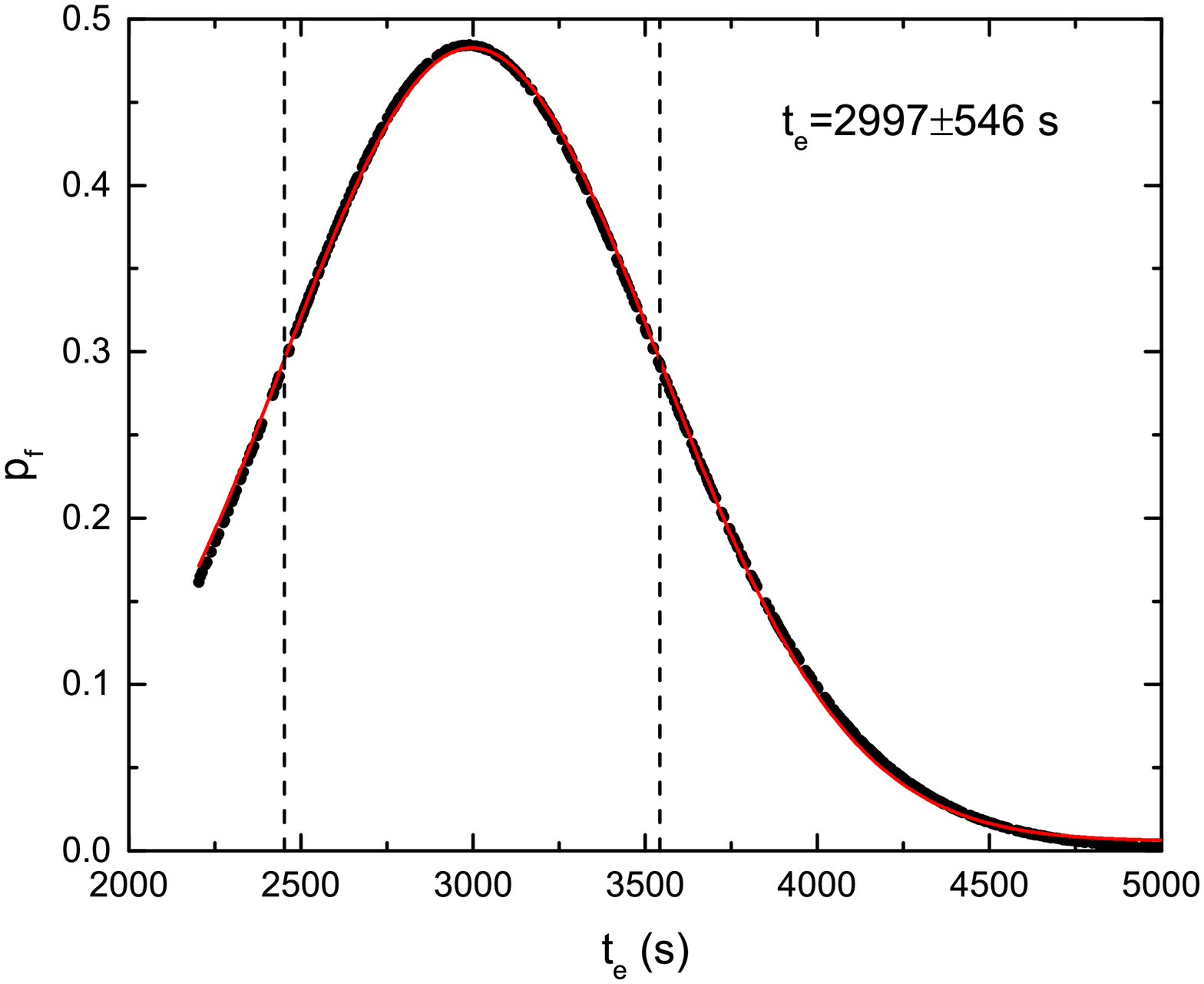}
\includegraphics[angle=0,scale=0.2]{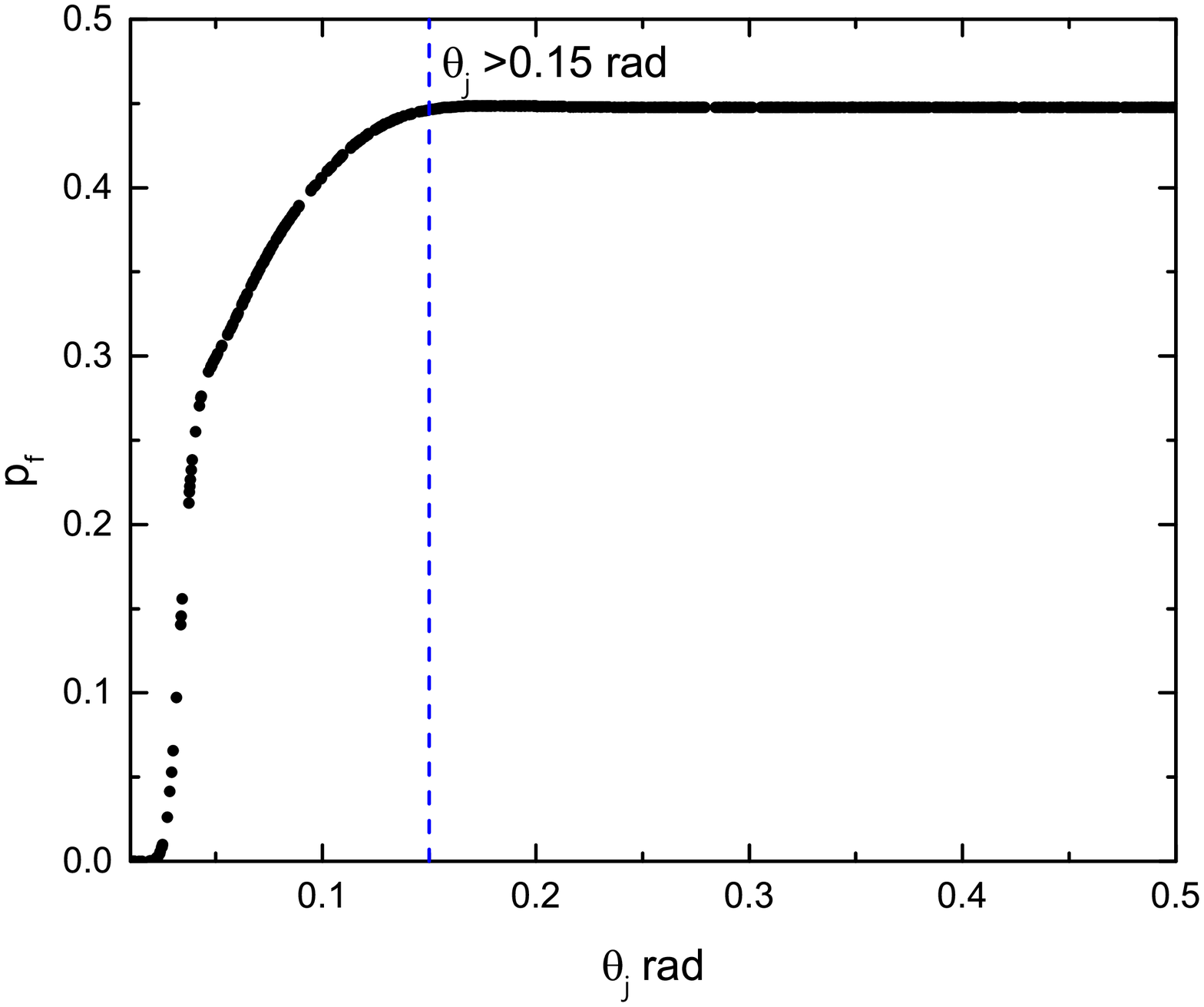}
\includegraphics[angle=0,scale=0.2]{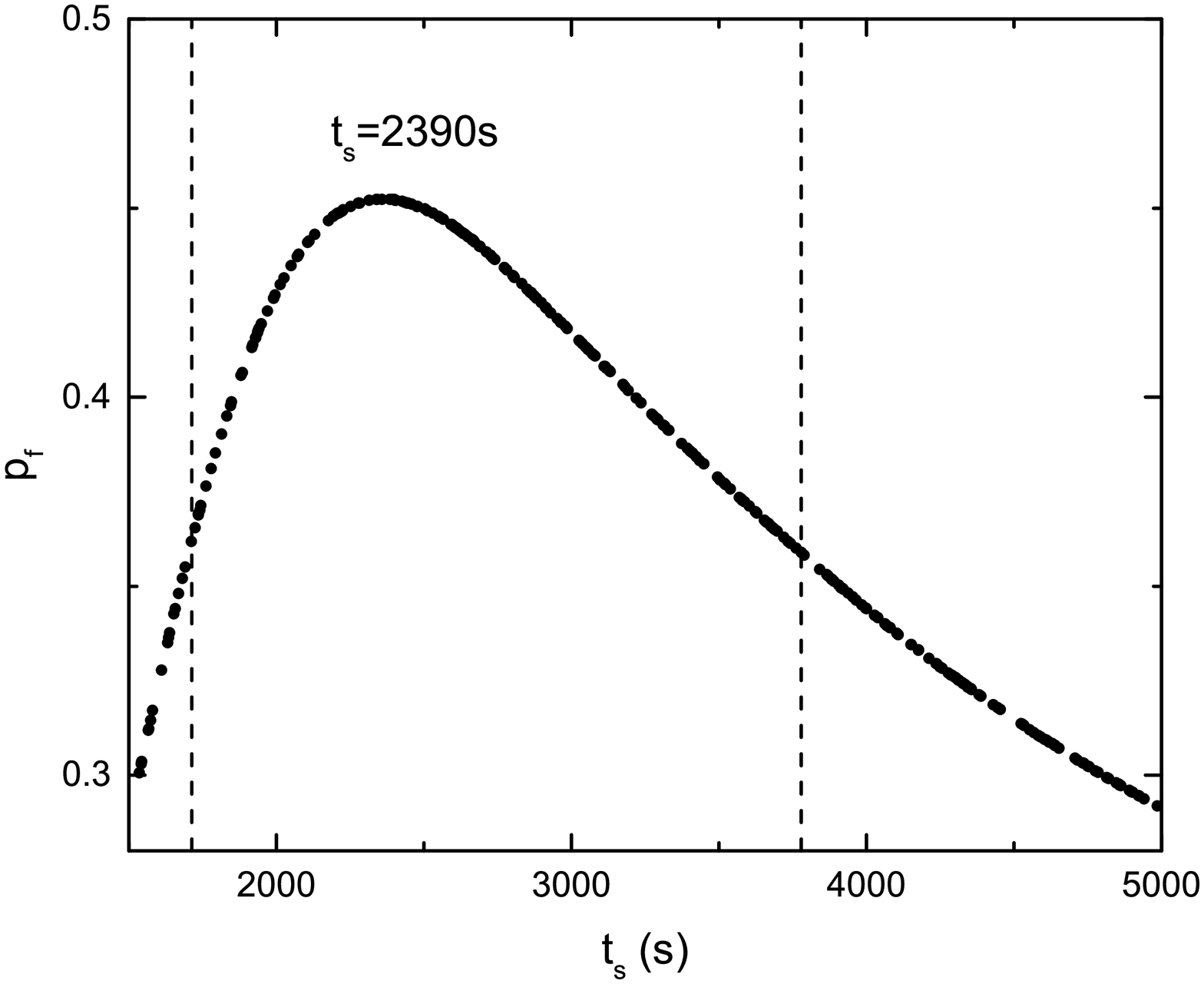}
\caption{Probability distributions of the forward shock adding the delayed energy injection model parameters along with our Gaussian
fits ({\em solid red lines}).The {\em dashed black vertical lines} mark the 1$\sigma$ standard deviations. Our fit gives a lower limit on $\theta_j$ only.}
\label{model_para}
\end{figure}

\begin{figure}[htbp]
 \centering
\includegraphics[angle=0,width=0.5\textwidth]{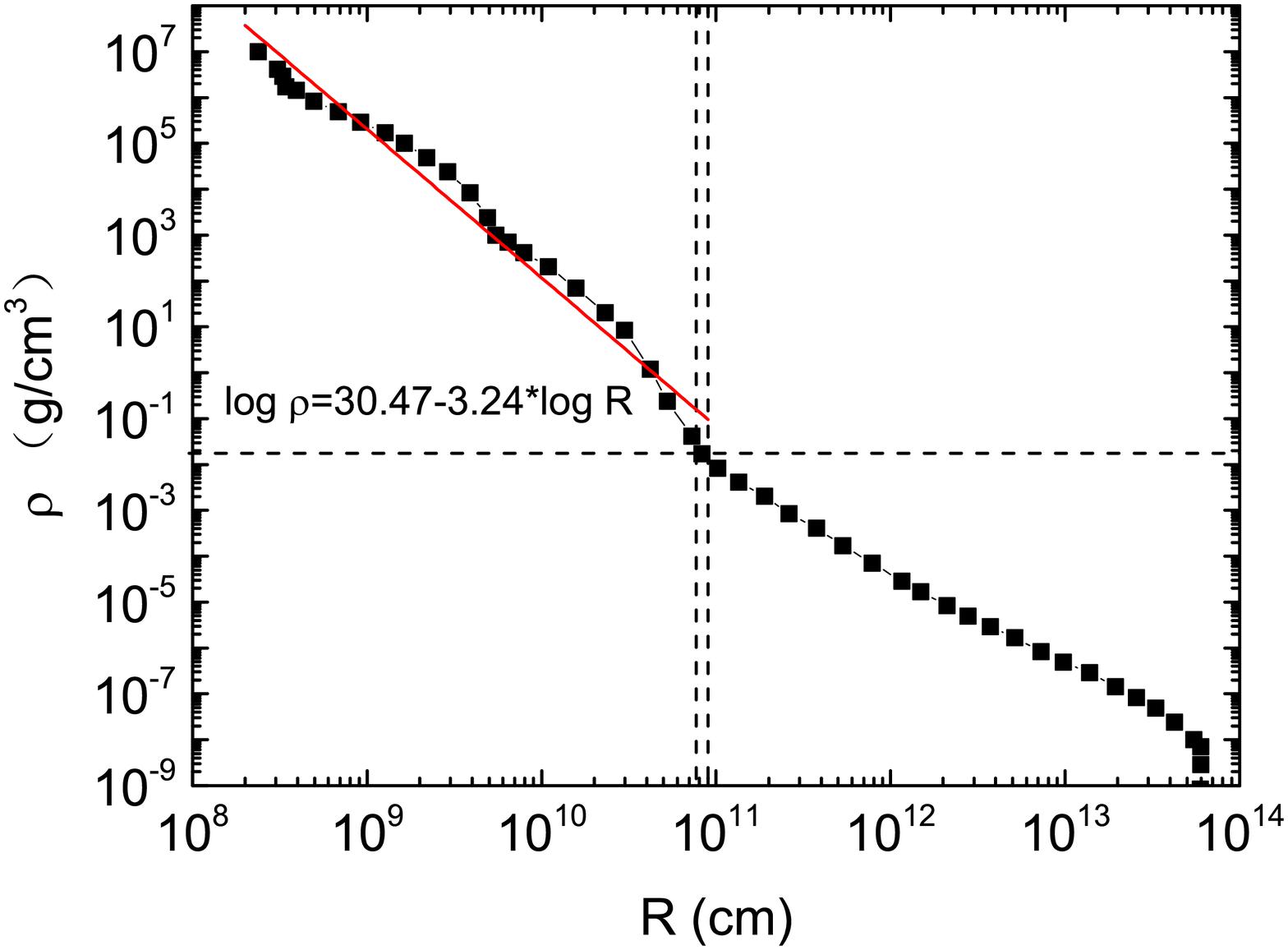}
\caption{Mass density profile as a function
of radius $R$ derived form simulations for a pre-supernova star with mass of 25M$_\odot$ (Woosley \& Weaver 1995). The vertical and horizonal {\em dashed lines} mark the radii and the corresponding density of fall-back materials for feeding the late accretion in this analysis.The {\em solid red line} is power-law fit to the density profile for $R<9\times 10^{10}$ cm.}
\label{mass_profile}
\end{figure}

\end{document}